\documentclass[fleqn,10pt]{wlscirep}
\title{Veiled symmetry of disordered Parity-Time lattices: protected PT-threshold and the fate of localization}

\author[1]{Andrew K. Harter}
\author[1]{Franck A. Onanga}
\author[1,*]{Yogesh N. Joglekar}

\affil[1]{Indiana University Purdue University Indianapolis (IUPUI), Indianapolis 46202, USA}
\affil[*]{yojoglek@iupui.edu}

\keywords{Parity-Time Symmetry, Disordered Systems, Localization}

\begin{abstract}
Open, non-equilibrium systems with balanced gain and loss, known as parity-time ($\mathcal{PT}$)-symmetric systems, exhibit properties that are absent in closed, isolated systems. A key property is the $\mathcal{PT}$-symmetry breaking transition, which occurs when the gain-loss strength, a measure of the openness of the system, exceeds the intrinsic energy-scale of the system. We analyze the fate of this transition in disordered lattices with non-Hermitian gain and loss potentials $\pm i\gamma$ at reflection-symmetric sites. Contrary to the popular belief, we show that the $\mathcal{PT}$-symmetric phase is protected in the presence of a correlated (periodic) disorder which leads to a positive $\mathcal{PT}$-symmetry breaking threshold. We uncover a veiled symmetry of such disordered systems that is instrumental for the said protection, and show that this symmetry leads to new localization behavior across the $\mathcal{PT}$-symmetry breaking transition. We elucidate the interplay between such localization and the $\mathcal{PT}$-symmetry breaking phenomena in disordered $\mathcal{PT}$-symmetric lattices, and support our conclusions with a beam-propagation-method analysis. Our theoretical predictions provide avenues for experimental realizations of $\mathcal{PT}$-symmetric systems with engineered disorder. 
\end{abstract}
\begin{document}

\flushbottom
\maketitle

\thispagestyle{empty}


\section*{Introduction}
Over the past decade, classical and quantum open systems in two categories have been intensely investigated for their non-equilibrium properties. The first category consists of systems that are in quasi-equilibrium and can be studied using linear response theory~\cite{kubo}. The second category has systems that are far removed from equilibrium~\cite{keldysh}, making perturbative methods inapplicable. Open systems with balanced gain and loss, called parity-time ($\mathcal{PT}$)-symmetric systems, straddle the two categories. In the quantum context, $\mathcal{PT}$-symmetric systems refer to those described by a non-Hermitian Hamiltonian $H_{PT}\neq H_{PT}^\dagger$ that is invariant under combined parity ($\mathcal{P}$) and time-reversal ($\mathcal{T}$) operations and leads to a non-unitary time evolution. The spectrum of $H_{PT}$ is purely real when the non-Hermiticity is small and becomes complex-conjugate pairs when it exceeds a threshold set by the Hermitian part of the Hamiltonian. This transition is called the $\mathcal{PT}$-symmetry breaking transition~\cite{bender0}. In the $\mathcal{PT}$-symmetric phase (real spectrum), the system is in a quasi-equilibrium state characterized by bounded, periodic oscillations in the system particle number. In the $\mathcal{PT}$-broken phase (complex spectrum), the system is far removed from equilibrium, and the particle number increases exponentially with time~\cite{review}. 

Two decades ago, $\mathcal{PT}$-symmetric Hamiltonians were first studied for continuum models on an infinite line~\cite{bender1,bender2,bender3}. The past five years, however, have made it clear that the experimentally relevant~\cite{exp1,exp2,exp3,exp4,exp5} ones are discrete lattice models~\cite{bendix,song,znojil1,avadh} or continuum models on a finite line~\cite{qma,serbyn,divergent}. For a one dimensional lattice with $N$ sites, the parity operator represents reflection about the lattice center, i.e., $\mathcal{P}_{mn}=\delta_{m\bar{n}}$ where $\bar{n}=N+1-m$ is the reflection-counterpart of site $n$. The time-reversal operator is given by complex conjugation, $\mathcal{T}=*$. A typical $\mathcal{PT}$-symmetric Hamiltonian consists of a Hermitian part $H_0$ that represents kinetic energy and a non-Hermitian part $\Gamma$ that represents balanced gain and loss. The $\mathcal{PT}$-symmetric nature of $H_0$ itself implies that its eigenfunctions are either symmetric or antisymmetric, ensures that the odd-order perturbative corrections from the gain-loss potential $\Gamma$ to the eigenenergies of $H_0$ vanish~\cite{moiseyev}, and thus leads to a positive $\mathcal{PT}$-symmetry breaking threshold. Discrete $\mathcal{PT}$-symmetric lattice Hamiltonians have been realized in coupled resonators~\cite{exp3,exp4,exp5} and coupled optical waveguides with balanced gain and loss~\cite{exp2}. 

Evanescently coupled optical waveguides are also an exceptional platform for simulating key quantum  phenomena~\cite{cow1} including Bloch oscillations~\cite{cow1half} and Anderson localization in one dimension due to arbitrarily weak disorder~\cite{cow2}. Although initially predicted in the condensed-matter context~\cite{bloch,anderson,g041,g042}, these phenomena have been thoroughly investigated in waveguide lattices because the Maxwell wave equation, under paraxial approximation, is isomorphic to the Schr\"{o}dinger equation for the wave-envelope function $|\psi(t)\rangle$~\cite{cow1}. In a sharp contrast with the nature-given lattices in condensed matter systems, waveguide lattices can be fabricated with a wide range of site-to-site tunneling amplitudes and on-site potentials; local or long-ranged "impurity" potentials; and on-site or tunneling disorder. This versatility has permitted the observation of disorder-induced localization, its insensitivity to the source of the disorder, as well as the signatures of the disorder-source in Hanbury-Brown Twiss correlations in disordered waveguide lattices~\cite{lahini1} and fibers~\cite{cow3}. 

What is the fate of a disordered $\mathcal{PT}$-symmetric system? In general, the $\mathcal{PT}$-symmetric phase is fragile in the sense that an arbitrarily weak disorder reduces the symmetry-breaking threshold to zero~\cite{bendix,moiseyev}. It does so because a random disorder does not preserve the symmetries of the underlying Hamiltonian. A straightforward way to salvage the fragile $\mathcal{PT}$-symmetric phase is to require a $\mathcal{PT}$-symmetric disorder~\cite{clintdisorder}. However, this approach imposes highly non-local correlations on the randomness and is therefore difficult to implement, even with an engineered disorder. Thus questions about localization and $\mathcal{PT}$-symmetry breaking in a disordered $\mathcal{PT}$-symmetric system appear moot~\cite{molina,konotop1}. 

In this report, we show that the $\mathcal{PT}$-symmetric phase in a disordered system is not always fragile, and that it is protected against random tunneling or on-site potential disorder if the disorder has specific periodicities. We elucidate an underlying symmetry that is critical for the said protection. We investigate the distribution of $\mathcal{PT}$-breaking threshold in such disordered systems and its dependence on the nature (tunneling or on-site potential) and the distribution (Gaussian, uniform, etc.) of disorder. In Hermitian disordered systems, disorder-averaged single particle properties, such as density of states and the localization profile, do not depend upon these details. Here, we show that the distribution of $\mathcal{PT}$-symmetry breaking threshold is sensitive to those disorder attributes. Our results demonstrate that a disordered $\mathcal{PT}$-symmetric system exhibits novel properties absent in its  Hermitian counterpart. 


\section*{Disordered lattice model}
Consider an $N$-site tight-binding lattice with gain and loss potentials $\pm i\gamma$ located at parity symmetric sites $m_0\leq N/2$ and $\bar{m}_0>N/2$ respectively; the lattice has open boundary conditions, meaning the first and the $N$th site has only one neighbor each. The distance between the gain and the loss sites, $d=\bar{m}_0-m_0$, ranges from $N-1$ to one (two) when $N$ is even (odd). The non-Hermitian, $\mathcal{PT}$-symmetric Hamiltonian for this lattice is given by $H_{PT}=H_0+\Gamma$ where 
\begin{eqnarray}
\label{eq:h0}
H_0 = & -J\sum_{k=1}^{N-1}\left(|k\rangle\langle k+1| +|k+1\rangle\langle k|\right) & =H_0^\dagger,\\
\label{eq:G}
\Gamma = & i\gamma\left( |m_0\rangle\langle m_0| -|\bar{m}_0\rangle\langle\bar{m}_0|\right) &=-\Gamma^\dagger.
\end{eqnarray}
$J>0$ is the constant tunneling amplitude that sets the energy-scale for the Hermitian Hamiltonian and $|k\rangle$ is a single-particle state localized at lattice site $k$. Since the Hamiltonian $H_{PT}$ commutes with the antilinear operator $\mathcal{PT}$, it follows that its spectrum is either purely real or consists of complex conjugate pairs~\cite{berry,ali}. The spectrum is real when $\gamma\leq\gamma_{PT}(m_0)$ where the $\gamma_{PT}(m_0)$ denotes the gain-location dependent $\mathcal{PT}$-symmetry breaking threshold. When $N$ is even, the threshold is maximum when the gain and loss potentials are nearest to each other or farthest away from each other, i.e., $\gamma_{PT}=J$ when $d=1$ and $d=N-1$. When $N$ is odd, $\gamma_{PT}\rightarrow J/2$ when $d=2$ and $\gamma_{PT}\rightarrow J$ when $d=N-1$. This unexpected robustness of the $\mathcal{PT}$-symmetry breaking threshold at the largest gain-loss distance is due to open boundary conditions~\cite{mark,derekring}. In the presence of a random, uncorrelated disorder, the threshold is suppressed to zero, i.e., $\gamma_{PT}=0$. In the following subsection, we show that introducing a periodic disorder alleviates this problem. 


\subsection*{$\mathcal{PT}$ phase diagram of a disordered lattice}
\label{subsec:ptphase}

\begin{figure}[]
\centering
\includegraphics[width=0.49\linewidth]{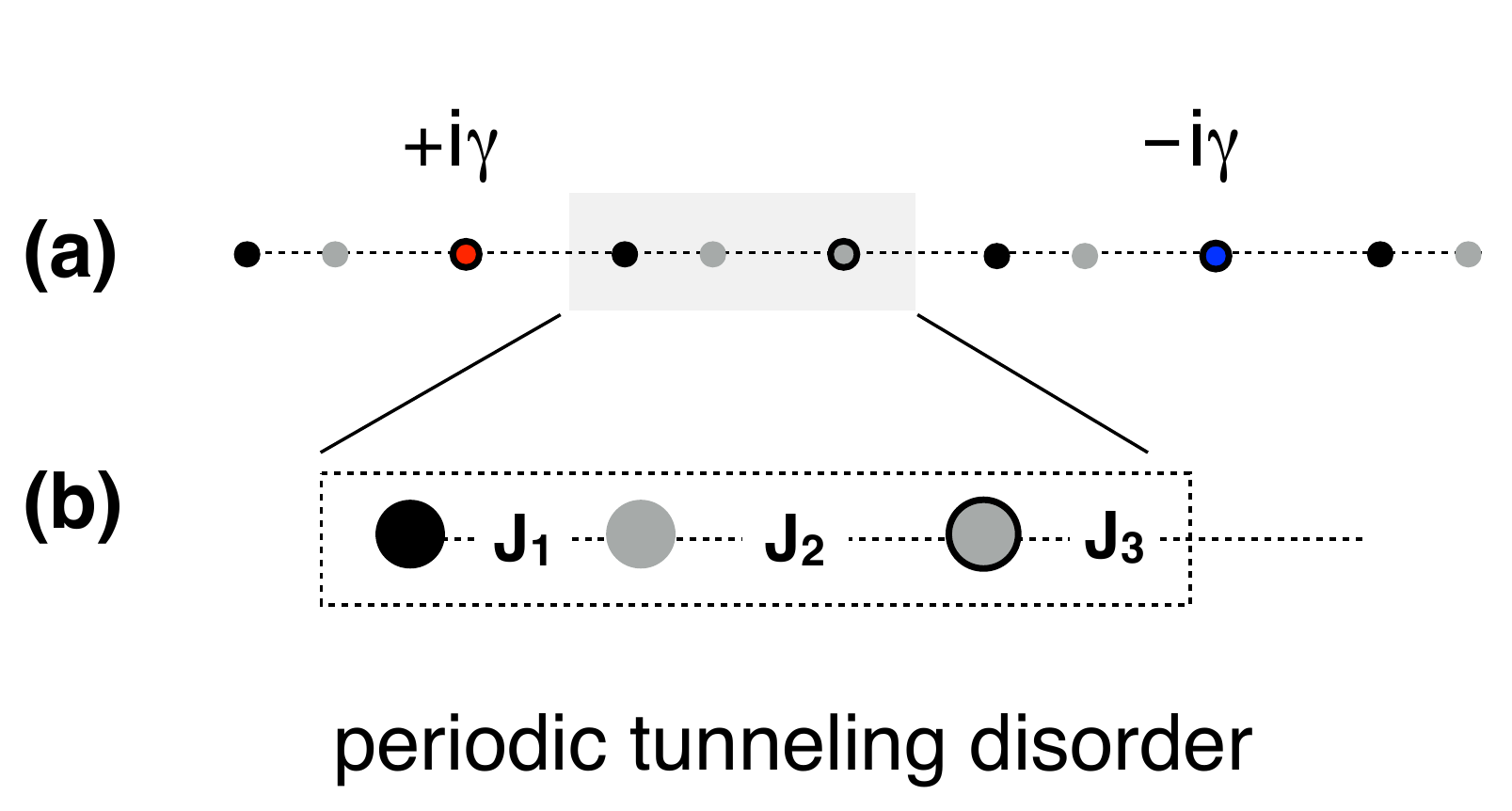}
\includegraphics[width=0.49\linewidth]{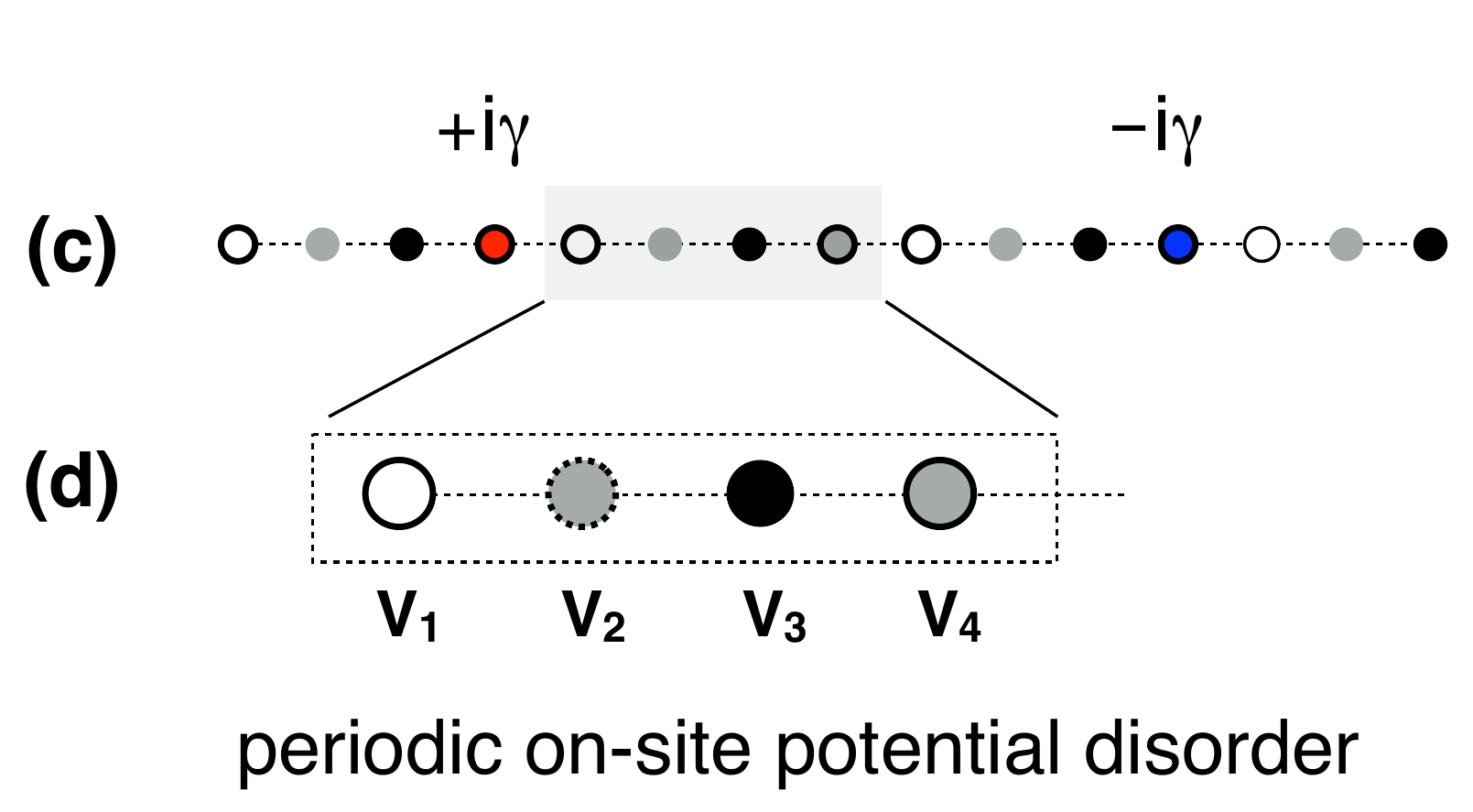}
\includegraphics[width=0.49\linewidth]{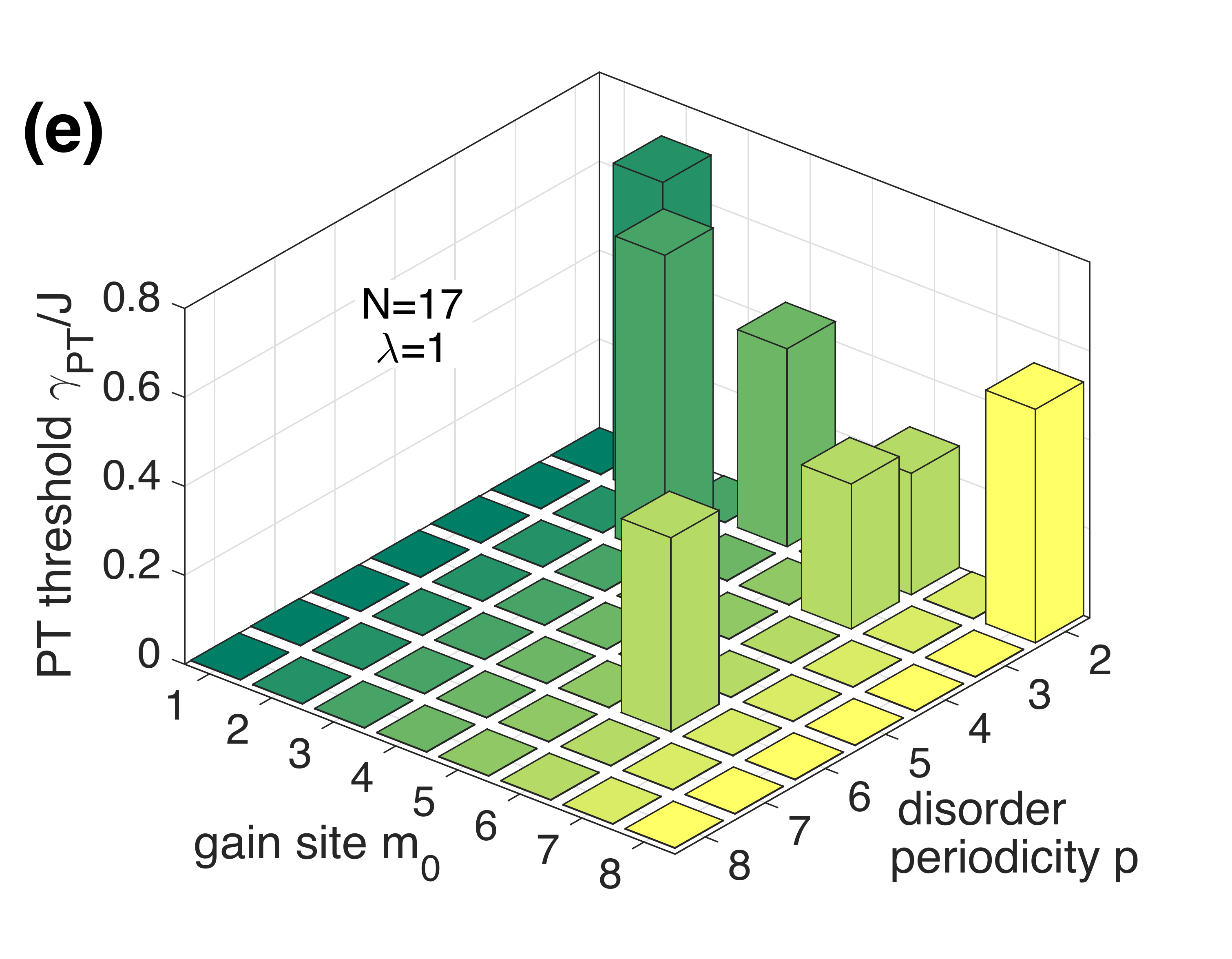}
\includegraphics[width=0.49\linewidth]{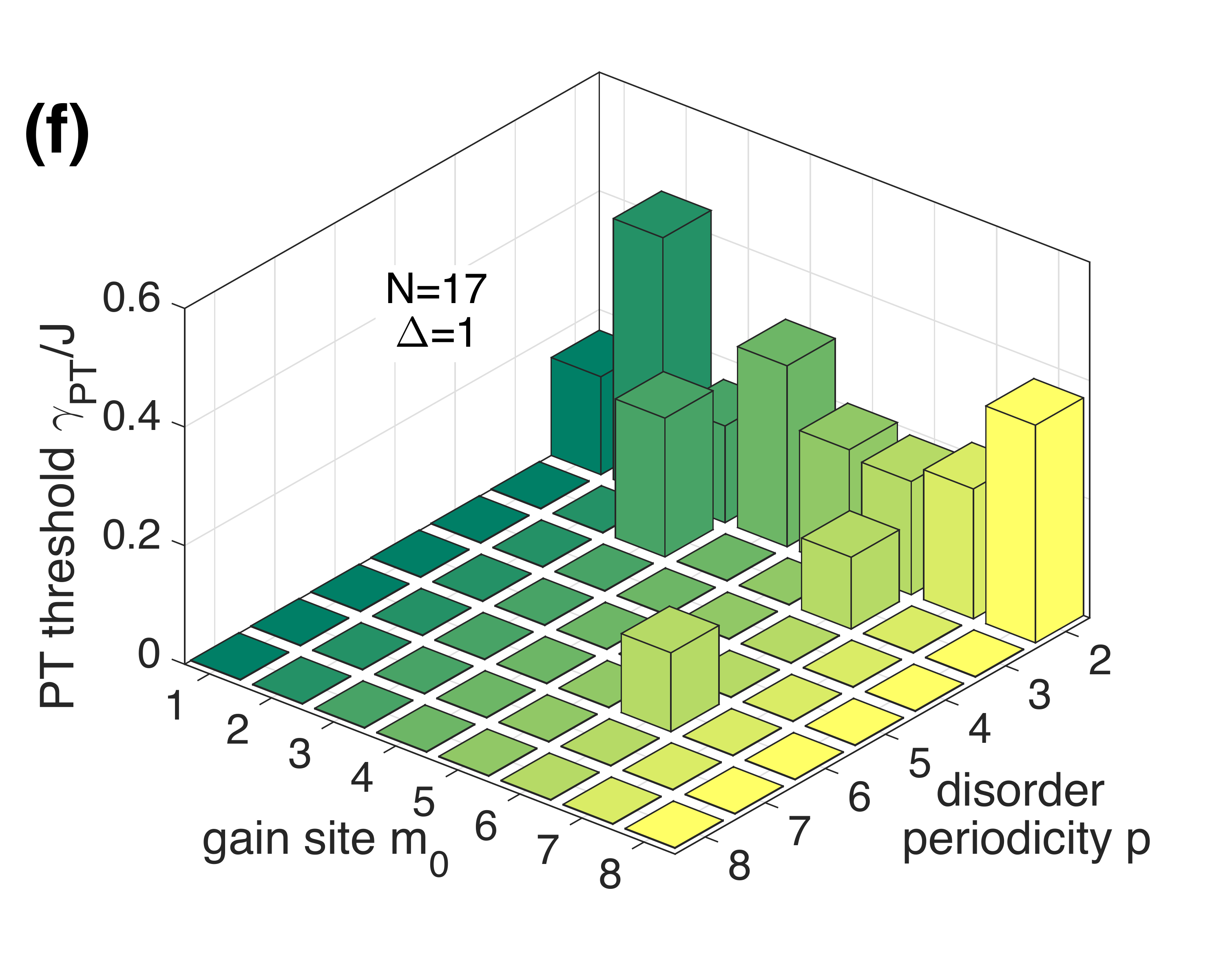}
\caption{Disordered $\mathcal{PT}$-symmetric lattices with open boundary conditions. {\bf (a)} An 11-site lattice with gain potential $+i\gamma$ at site $m_0=3$ and random periodic tunneling $J_k=J(1+\lambda r_k)$. Here $\lambda=1$ is the strength of the disorder and $\{r_1,\ldots,r_p\}$ are $p$ random numbers with zero mean and unit variance. {\bf(b)} The tunneling disorder has period $p=3$. {\bf (c)} A 15-site lattice with uniform tunneling, $m_0=4$, and random on-site potentials $V_k=J\Delta r_k$ with $\Delta=1$. {\bf (d)} The potential disorder has period $p=4$. {\bf (e)} $\mathcal{PT}$-symmetry breaking threshold $\gamma_{PT}(m_0)$ as a function of gain site $m_0\leq N/2$ and tunneling disorder period $p\leq N/2$ shows that $\gamma_{PT}>0$ when $N+1=0\mod p$ and $m_0=0\mod p$; it is zero otherwise. {\bf(f)} Results for on-site disorder show the same behavior except at $p=2$ for an odd $N$. Then the on-site disorder is $\mathcal{PT}$ symmetric, and $\gamma_{PT}(m_0)>0$ for all $m_0$. These results imply that the positive $\mathcal{PT}$-breaking threshold of a uniform lattice is protected from a correlated disorder under the right circumstances.} 
\label{fig:ptphase}
\end{figure}

We consider two classes of Hermitian disorders, one in the tunneling amplitude and the second in the on-site potential, each with lattice period $p$, 
\begin{eqnarray}
\label{eq:vod}
V_T & = & J\lambda\sum_{k=1}^{N-1} r_k \left( |k\rangle\langle k+1| +|k+1\rangle\langle k|\right),\\
\label{eq:vd}
V_O & = & J\Delta\sum_{k=1}^N r_k |k\rangle\langle k|.
\end{eqnarray}
The dimensionless numbers $\lambda\geq 0$ and $\Delta\geq 0$ represent the strength of tunneling and on-site disorder respectively, $\{r_1,\ldots,r_p\}$ are independent, identically distributed (i.i.d.) random numbers with zero mean and unit variance, and the periodic nature of disorder implies that  $r_{k'}=r_k$ if $k'-k=0\mod p$. Figure~\ref{fig:ptphase} (a)-(b) show the schematic of a disordered lattice with $N=11$ sites and gain potential $i\gamma$ at site $m_0=3$. The tunneling disorder $V_T$ has period $p=3$, and the three independent, random tunnelings are given by $J_k=J(1+\lambda r_k)$. Figure~\ref{fig:ptphase} (c)-(d) show an on-site-potential disordered lattice with $N=15$ sites, gain potential at site $m_0=4$, and disorder period $p=4$; the four independent, random potentials are given by $V_k=J\Delta r_k$. Note that the periodic disorder potential in each case is not $\mathcal{PT}$-symmetric, i.e., $[\mathcal{PT},V_{T,O}]\neq 0$. Therefore, conventional wisdom suggests that the $\mathcal{PT}$-symmetry breaking threshold in each case will be zero. 

Figure~\ref{fig:ptphase} (e) shows the numerically obtained threshold $\gamma_{PT}(m_0,p)$ for an $N=17$ lattice with tunneling disorder strength $\lambda=1$. The key features of the threshold phase diagram are as follows. It is nonzero only when $N+1=0\mod p$ and $m_0=0\mod p$. Thus, when $p=2$ the threshold is nonzero only when $m_0$ is even, for $p=3$ it is nonzero for $m_0=\{3,6\}$, and for $p=6$, it is nonzero only when $m_0=6$. It is identically zero for periods $p=\{4,5,7,8\}$ for any gain-site location $m_0$. These results, obtained for a particular realization of the tunneling disorder, are generic. They show that a tunneling disorder with appropriately chosen period $p$ and gain locations $m_0$ leads to a positive $\mathcal{PT}$-symmetry breaking threshold with values comparable to that of a clean system, $\gamma_{PT}\sim J$. 

Figure~\ref{fig:ptphase} (f) shows the corresponding threshold results for an on-site disorder with strength $\Delta=1$. The salient features of the phase diagram are the same: $\gamma_{PT}>0$ when $N+1=0\mod p$ and $m_0=0\mod p$. Thus, periodicities $p=\{2,3,6\}$ have a positive threshold for appropriate gain locations, while $\gamma_{PT}=0$ for all other disorder periods. In addition, when $p=2$  (on-site, dimer disorder), the symmetry breaking threshold is nonzero for odd values of gain location as well. This is the only qualitative difference between the threshold results for tunneling vs. on-site disorders. It arises because for an odd $N$ and $p=2$, the on-site disorder is always $\mathcal{PT}$-symmetric, i.e., $[\mathcal{PT},V_O]=0$. For an even lattice, both tunneling and on-site dimer disorders have $\gamma_{PT}>0$ only when the gain potential site is even. 

Results in Figure~\ref{fig:ptphase} (e)-(f) are surprising because {\it they show that the symmetry breaking threshold is robust against disorders that are not parity symmetric}~\cite{aah}. They hint at the existence of another (antilinear) symmetry~\cite{berry,ali} that protects the threshold. In the next subsection, we uncover this veiled symmetry and discuss its consequences.


\subsection*{The $\Pi$-operator and a veiled symmetry}
\label{sec:hidden}

The tunneling Hamiltonian of a uniform lattice can be expressed as $H_0= U D U^{\dagger}$ where $D_{\alpha\beta}=\epsilon_\alpha\delta_{\alpha\beta}$ is a diagonal matrix with eigenvalues $\epsilon_\alpha=-2J\cos p_\alpha$, the unitary matrix with corresponding eigenfunctions has entries $U_{m\alpha}=\sqrt{2/(N+1)}\sin(p_\alpha m)$, and $p_\alpha=\pi\alpha/(N+1)$ are the quasimomenta consistent with open boundary conditions. The spectrum of $H_0$ is particle-hole symmetric, i.e., its eigenvalues satisfy $\epsilon_{\bar{\alpha}}=-\epsilon_\alpha$ where $\bar{\alpha}=N+1-\alpha$. The eigenfunctions of $H_0$ satisfy $U_{\bar{m}\alpha}=(-1)^{\alpha-1}U_{m\alpha}$ and $U_{m\bar{\alpha}}=(-1)^{m-1}U_{m\alpha}$. The first equation implies that the eigenfunctions are either symmetric or antisymmetric; the second equation implies that the particle and hole eigenfunctions, i.e. eigenfunctions with opposite energies, are related by a staggered $\pi$-phase. 

For a given $H_0$, we can generate a family of "parity" operators $P= USU^\dagger$ where $S=\mathrm{diag}(\pm,\ldots,\pm 1)$ is a diagonal matrix with randomly chosen entries $\pm 1$; there are $2^{N-1}$ such distinct operators. When $S=1_N$ the resultant operator is the identity and when $S_{kk'}=(-1)^{k-1}\delta_{kk'}={\mathcal S}$, the resultant parity operator is $\Pi=U{\mathcal S}U^\dagger=\mathcal{P}$. In the site-space representation, both matrices are sparse. For a random string in $S$, the resultant "parity" operator is not a sparse matrix. This procedure is generalized to the case of a disordered Hermitian Hamiltonian $H=H_0+V(\lambda,\Delta)$ and leads to a family of $2^{N-1}$ {\it disorder-dependent operators} $P(\lambda,\Delta)$. It is easy to show that $P(\lambda,\Delta)$ is Hermitian, $P^2=1$, $P\mathcal{T}=\mathcal{T}P$, and $P(\lambda,\Delta)$ commutes with the disordered Hamiltonian $H(\lambda,\Delta)$. Note that the special parity operator $\Pi(\lambda,\Delta)=U(\lambda,\Delta){\tt S}U(\lambda,\Delta)^\dagger$ depends on the disorder and is not equal to the parity (reflection) operator on the lattice, i.e., $\Pi(\lambda,\Delta)\neq\mathcal{P}$. 

\begin{figure}[htpb]
\centering
\includegraphics[width=0.49\linewidth]{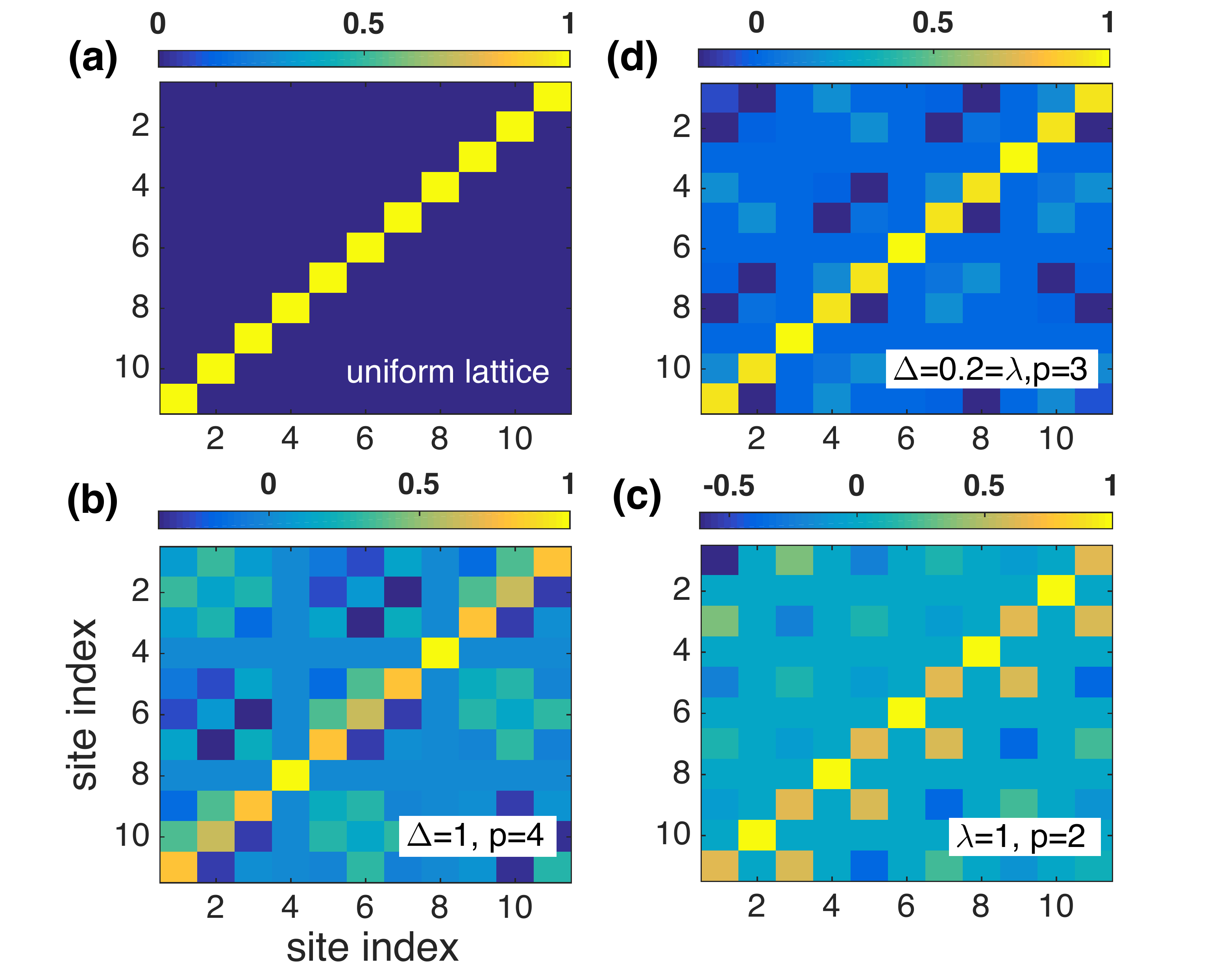}
\includegraphics[width=0.50\linewidth]{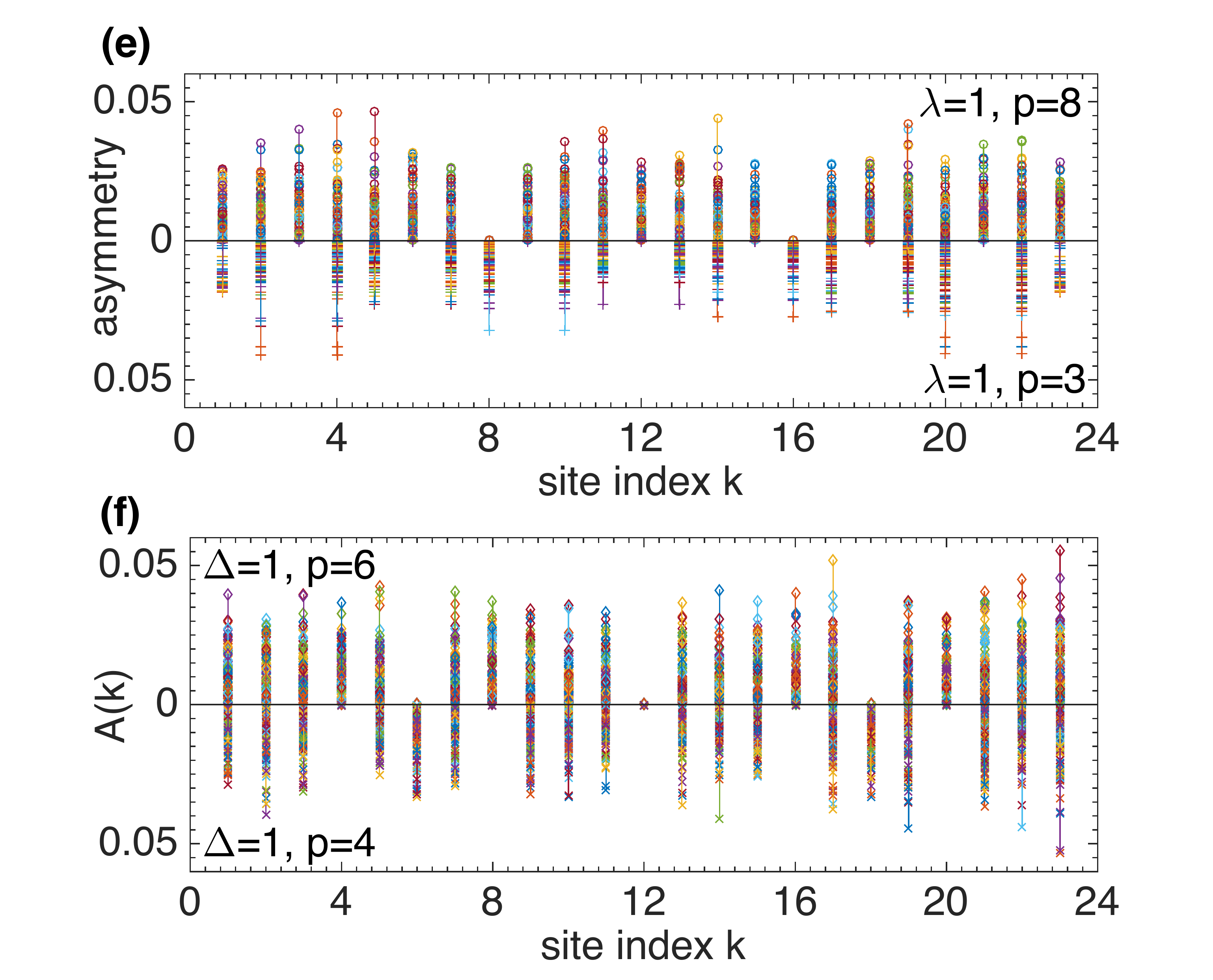}
\caption{Veiled symmetry of a disordered lattice. {\bf (a)} For a uniform lattice, the operator $\Pi=\mathcal{P}$ is given by $\Pi_{kk'}=\delta_{k\bar{k}}$. Typical parity operators $\Pi$ for an $N=11$ site lattice with different disorder strengths $\lambda,\Delta$ and periods $p$ are shown in {\bf (b)}-{\bf(d)}. In each case $\Pi_{k\bar{k}}=1$ if and only if $k=0\mod p$ and $N+1=0\mod p$. {\bf (e)} The site-dependent asymmetry functions $A(k)$ for an $N=23$ site lattice with tunneling disorder. The disorder strength is $\lambda=1$ and the number of disorder realizations is $M=100$. The asymmetry vanishes only if the disorder period $p$ satisfies $N+1=0\mod p$ and only on sites $k$ that are integer multiples of the disorder period $p$. {\bf(f)} Results for an on-site disorder with strength $\Delta=1$ show the same quantitative trend. This veiled symmetry of the eigenfunctions of a disordered lattice is instrumental to the positive $\mathcal{PT}$-symmetry breaking thresholds in Figure.~\ref{fig:ptphase}.}
\label{fig:parities}
\end{figure}

Figure~\ref{fig:parities} (a)-(d) show typical features of the $\Pi$ operator in the site basis. For a uniform lattice, panel (a), the $\Pi$ operator is the same as the lattice reflection operator. In the presence of disorder with period $p$, $\Pi$ is not a sparse matrix and satisfies $\Pi_{k\bar{k}}=1$ if and only if both $N+1=0\mod p$ and $k=0\mod p$ hold true. These results are generic and apply for on-site potential disorder, panel (b); tunneling disorder, panel (c); or a combination of the two, panel (d). In all cases, $[H,\Pi\mathcal{T}]=0$. A positive symmetry-breaking threshold, then, is possible if and only if the antilinear operator $\Pi\mathcal{T}$ also commutes with the gain-loss potential $\Gamma$, eq.(\ref{eq:G}). It is straightforward, albeit tedious, to verify that it is so only when $N+1=0\mod p$ and $m_0=0\mod p$. 

An insight into the vanishing commutator, $[\Gamma(m_0),\Pi\mathcal{T}]=0$, is offered by the effect of periodic disorder on the eigenfunctions of the uniform lattice.  When the disorder is zero, the eigenfunctions $U_{m\alpha}$ are symmetric or antisymmetric, i.e., $U_{\bar{m}\alpha}=(-1)^{\alpha-1}U_{m\alpha}$. They have equal weights on reflection-symmetric sites $m$ and $\bar{m}$ for all $m$. This property ensures that odd-order perturbative corrections to the eigenvalues $\epsilon_\alpha$ due to the gain-loss potential $\pm i\gamma$ vanish, and leads to a positive $\mathcal{PT}$ breaking threshold~\cite{moiseyev}. Are the eigenfunctions of the disordered Hamiltonian also reflection symmetric? To quantify this property, we define a site-dependent asymmetry function
\begin{equation}
\label{eq:asymmetry}
A(k)=\sum_{\alpha=1}^N \left|U_{\bar{k}\alpha}(\lambda,\Delta)+(-1)^\alpha U_{k\alpha}(\lambda,\Delta)\right|.
\end{equation}
It follows that $A\geq 0$ in general and for a uniform lattice, $A(k)=0$ for all $k$. The asymmetry functions $A(k)$ for $M=100$ tunneling disorder realizations on an $N=23$ site lattice are shown in Figure~\ref{fig:parities} (e). When the disorder period is $p=8$ (top panel), $A(k)=0$ only at sites $k=\{8,16\}$,  whereas when $p=3$ (bottom panel) the function vanishes exactly when $k=0\mod 3$. Figure~\ref{fig:parities} (f) has the corresponding results for an on-site disorder. Once again, we see that $A(k)=0$ if and only if $k=0\mod p$. The asymmetry function is nonzero everywhere if either of the two constraints, $N+1=0\mod p$ and $k=0\mod p$, is not satisfied. 

Results in Figure~\ref{fig:parities} (e)-(f) show that the disordered eigenfunctions $U_{m\alpha}(\lambda,\Delta)$ are neither symmetric nor antisymmetric, but, when restricted to a specific set of sites, they show these symmetries~\cite{aah}. It follows that $[\Gamma(m_0),\Pi\mathcal{T}]=0$ if and only if $m_0=0\mod p$ and $N+1=0\mod p$. Thus, although the Hamiltonian $H_{PT}(\lambda,\Delta)=H(\lambda,\Delta)+\Gamma$ is not $\mathcal{PT}$-symmetric, it is $\Pi\mathcal{T}$-symmetric under these constraints. This veiled antilinear symmetry of the eigenfunctions of $H$ gives rise to the positive $\mathcal{PT}$ breaking thresholds seen in Figure~\ref{fig:ptphase}. 


\subsection*{Disorder induced $\mathcal{PT}$ threshold distribution and localization}
\label{sec:distribution}

Disordered models with positive $\mathcal{PT}$-symmetry breaking thresholds prompt a number of questions. How does the probability distribution function of the $\mathcal{PT}$-breaking threshold $PDF(\gamma_{PT})$ depend on the strength of the disorder? Does it depend on the distribution of the disorder? Is it different for on-site and tunneling disorders? What is the fate of localization in $\mathcal{PT}$-symmetric systems? These questions are addressed in the following paragraphs. 

\begin{figure}[ht]
\centering
\includegraphics[width=0.49\linewidth]{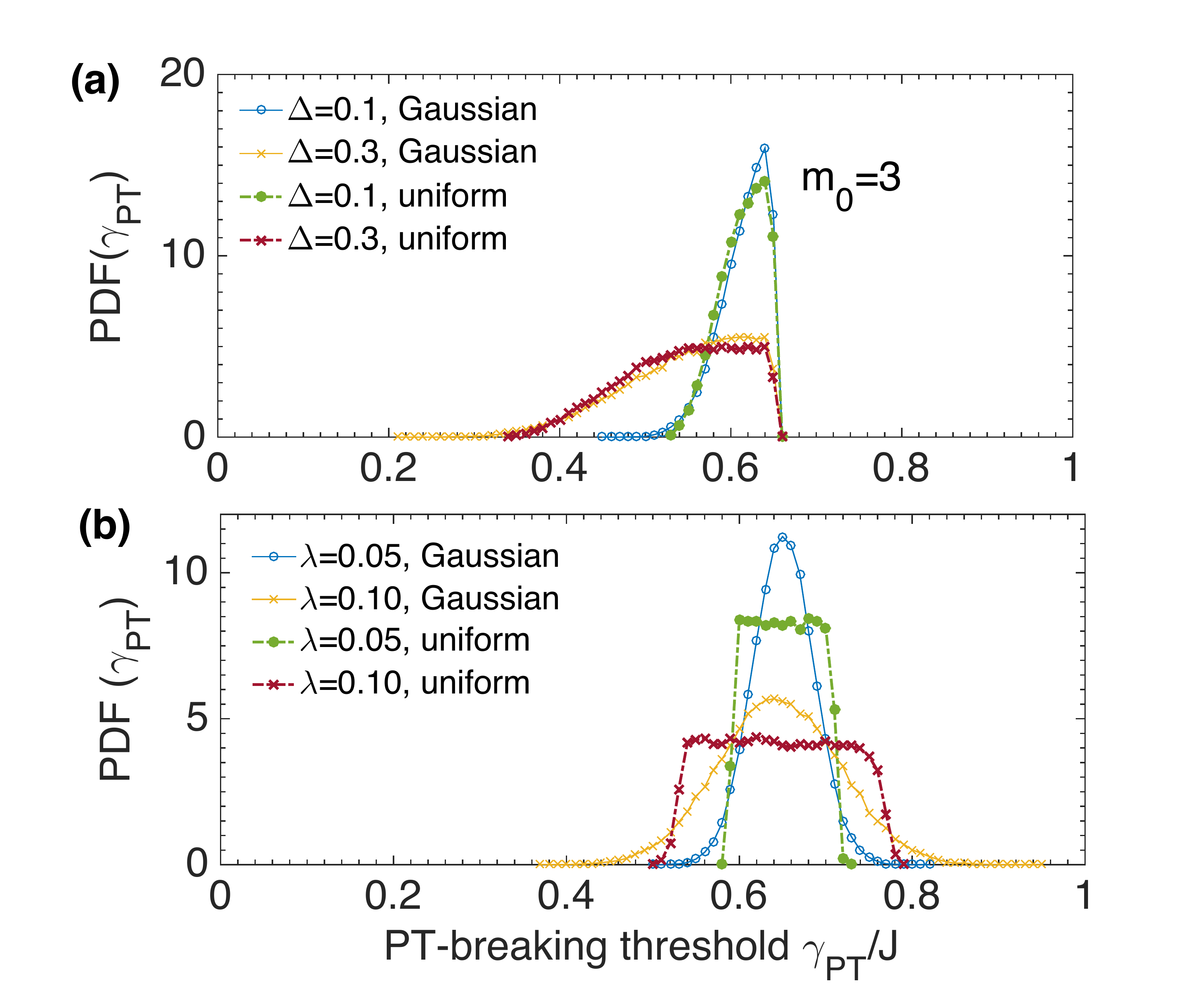}
\includegraphics[width=0.50\linewidth]{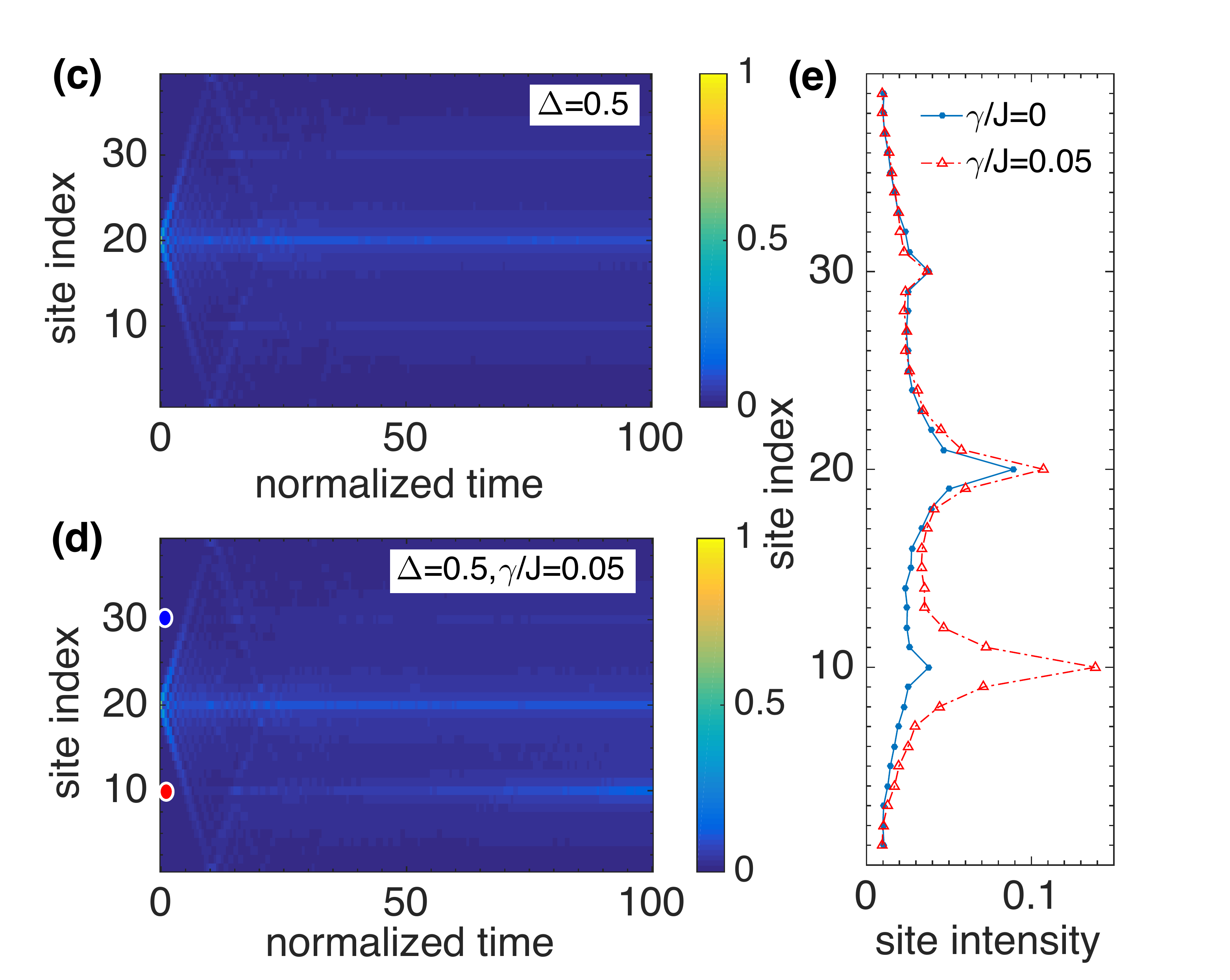}
\caption{$\mathcal{PT}$-symmetry breaking threshold distribution $PDF(\gamma_{PT})$ for the gain potential at site $m_0=3$, in an $N=17$ site lattice with disorder period $p=3$. {\bf(a)} For an on-site potential disorder, the threshold distribution $PDF(\gamma_{PT})$ broadens as disorder strength $\Delta$ increases and it is independent of the disorder distribution, Gaussian or uniform. {\bf(b)} For the tunneling disorder, the threshold distribution $PDF(\gamma_{PT})$ mimics the disorder distribution, giving different results for a Gaussian disorder and the uniform disorder. {\bf(c)} Localization in an $N=39$ site lattice, with on-site disorder period $p=10$, the initial state at the center of the lattice, and $M=1000$ disorder realizations. When $\gamma=0$, the disorder-averaged intensity $I_d(k,t)$ shows satellite peaks at $k=k_0\mod p$ in addition to the usual peak at the initial site $k_0=20$. {\bf(d)} when the gain-potential is turned on, $\gamma/J=0.05$, intensity weight at the gain site $m_0=10$ increases with time. {\bf(e)} Intensity profile $I_d(k,t)$ at time $Jt=100$ shows that the increase in the intensity at the gain-site when $\gamma>0$ (red open triangles) does not come at the expense of the intensity at other sites, but instead from the non-unitary time evolution.}
\label{fig:loc}
\end{figure}

Figure~\ref{fig:loc} shows the threshold distributions $PDF(\gamma_{PT})$ in the presence of on-site potential disorder, panel (a), and tunneling disorder, panel (b). The results are for the $\mathcal{PT}$-symmetry breaking threshold at gain site $m_0=3$ in an $N=17$ lattice, obtained by using $M=5\mathrm{x}10^4$ realizations of disorder with period $p=3$. The horizontal axis in each panel is the dimensionless threshold $\gamma_{PT}/J$. Panel (a) shows that as the on-site disorder strength $\Delta$ increases, the threshold distribution $PDF(\gamma_{PT})$ becomes broader, and skewed towards values smaller than its clean-limit value. In addition, $PDF(\gamma_{PT})$ is independent of the disorder distribution, i.e., it is the same whether the random, periodic potential is drawn from a Guassian distribution with zero mean and variance $\Delta$ (blue open circles, yellow crosses) or a uniform distribution with the same mean and variance (green filled circles, red crosses). Qualitatively similar results are obtained for other lattice sizes $N$, disorder periods $p$, and gain potential locations $m_0$ as long as they satisfy the criteria $N+1=0\mod p$ and $m_0=0\mod p$. These results are consistent with what we would expect. Introducing disorder suppresses the $\mathcal{PT}$-breaking threshold and the threshold distribution $PDF(\gamma_{PT})$ - a single particle property - is independent of the underlying disorder distribution~\cite{rmt}. 

Figure~\ref{fig:loc} (b) shows that these expectations are rather simplistic. For a Gaussian tunneling disorder (blue open circles, yellow crosses), $PDF(\gamma_{PT})$ is a bell shaped distribution centered about its clean-limit value. It becomes broader when the tunneling disorder strength $\lambda$ is increased, and its center shifts towards the origin. For a uniform disorder (green filled circles, red crosses), we find that $PDF(\gamma_{PT})$ is now a flat-top distribution approximately centered about its clean-limit value. We obtain qualitatively similar results for other lattice sizes, gain locations, and disorder periodicities that lead to a positive clean-limit  threshold. These results are remarkable because {\it for a tunneling disorder, the threshold distribution $PDF(\gamma_{PT})$ mimics the disorder distribution and is not universal}. 

In one-dimensional Hermitian systems, a random disorder exponentially localizes all states. In transport experiments, this localization is inferred from a scaling analysis of the resistivity in the presence of disorder~\cite{g041,g042}. In optical-waveguide realizations of a Hermitian disordered lattice, it is manifest by a disorder-averaged intensity profile that, after an initial ballistic expansion, develops a steady-state value~\cite{cow1,cow2,lahini1}. For an initial state on site $k_0$, the disorder-averaged intensity profile $I_d(k,t)=|\langle k|\psi(t)\rangle|^2_d$ is symmetrically and exponentially localized around that site. Here the subscript $d$ denotes averaging over different disorder realizations, $|\psi(t)\rangle=G(t)|\psi(0)\rangle$, and the time-evolution operator is $G(t)=\exp(-iHt)$ where we have used $\hbar=1$. In the Hermitian case, $\gamma=0$, the time-evolution operator is unitary, i.e., $G(t)^\dagger G(t)=1$. Therefore, the total disorder-averaged intensity is constant at each time, $\sum_k I_d(k,t)=1$. 

In the $\mathcal{PT}$-symmetric disordered case, there are two distinct scenarios. If the gain potential strength is smaller than the minimum threshold value, i.e.,  $\gamma<\gamma_{\min}=\min_{\gamma}\{PDF(\gamma_{PT})>0\}$, the system is in the $\mathcal{PT}$-symmetric phase for each disorder realization. Therefore, its non-unitary time evolution has bounded intensity oscillations and at long times $Jt\gg1$, it leads to a quasi steady-state intensity profile $I_d(k)$ with constant total intensity $\sum_k I_d(k)>1$~\cite{review,clintdisorder}. When $\gamma>\gamma_{\min}$, for a fraction of the $M\gg1$ disorder realizations, the system is in the $\mathcal{PT}$-broken phase where the total intensity increases exponentially with time as does the intensity in the neighborhood of the gain site $m_0$. As a result, the disorder-averaged intensity $I_d(k,t)$ develops a peak at the gain site $m_0$ whose weight increases with time. We note that in this regime, the intensity $I_d(k,t)$ does not reach a steady state value~\cite{review,molina,konotop1}.    

Figure~\ref{fig:loc} (c)-(e) encapsulates the effects of correlated disorder on the disorder-averaged site- and time-dependent intensity $I_d(k,t)$. The results are for an $N=39$ site lattice with on-site potential disorder with periodicity $p=10$, number of disorder realizations $M=10^3$, and an initial state localized at the center of the lattice, $k_0=20$. Panel (c) shows the disorder-averaged intensity $I_d(k,t)$ for the Hermitian case, $\gamma=0$. A periodic disorder leads to a steady-state profile $I_d(k)$ that is exponentially localized about the initial site $k_0=20$, along with satellite peaks at sites $k=k_0\mod p=\{10,30\}$. These satellite peaks are signatures of extended states that exist in one-dimensional systems with periodic disorder~\cite{pd1,pd2}. As the disorder strength $\Delta$ is increased, the peak intensity of the satellites decreases. We remind the reader that when the disorder is purely random, and not periodic, these satellite peaks are absent. 

Panel (d) shows corresponding results for a disordered $\mathcal{PT}$-symmetric system with gain potential of strength $\gamma/J=0.05$ at site $m_0=p=10$ (red filled circle); the corresponding loss potential $-i\gamma$ at site $\bar{m}_0=30$ is also shown (blue filled circle). We see that in addition to the hermitian localization peaks at sites $k=k_0\mod p$, a new peak emerges at the gain location. It arises because a disordered system with $\gamma/J=0.05$ is, sometimes, in the broken $\mathcal{PT}$-symmetric phase. Panel (e) shows the disorder-averaged site-intensity profile $I_d(k,t)$ at time $Jt=100$. In the Hermitian case, it shows localization peaks at the initial site $k_0=20$ and satellite peaks at sites $k=k_0\mod p=\{10,30\}$ (blue filled circles). In the $\mathcal{PT}$-symmetric case, the intensity values are essentially unchanged except in the vicinity of the gain site, where the it has increased by a factor of five (red open triangles). This interplay between the disorder-induced localization and the broken $\mathcal{PT}$-symmetry induced localization occurs even if there is no disorder-induced peak at $m_0$  in the Hermitian limit.


\subsection*{Beam propagation method analysis}
\label{sec:bpm}
The results for a finite $\mathcal{PT}$-symmetry breaking threshold in disordered lattices presented in Figure~\ref{fig:ptphase} are based on a tight-binding approximation. In the experimental realizations of such lattices, however, a "site" has a transverse spatial extent, and the tunneling Hamiltonian  in Eq.(\ref{eq:h0}) represents a site-discretized version of the spatial second derivative in the continuum Schr\"{o}dinger (or Maxwell) equation. Therefore, to test that our predictions are not artifacts of the lattice approximation, we obtain the time-evolution of the wave function $\psi(x,t)$ in a waveguide array with realistic parameters~\cite{bpm3} via the beam propagation method (BPM)~\cite{bpm1,bpm2}. The continuum Sch\"{o}dinger equation is given by $i\partial_t\psi=-\partial^2_x\psi/2m +V(x)\psi$. Here, the effective mass is $m=k_0n_0^2/c$, the potential is given by $V(x)=ck_0\left[ 1- n(x)^2/n_0^2\right]$, $n_0$ is the cladding index of refraction, $c$ is the speed of light in vacuum, $n(x)$ is the position-dependent index of refraction in the waveguide array, and $k_0=2\pi/\lambda$ is the wave number of the rapidly varying part of the electric field $E(x,z,t)=\exp\left[ik_0z-(ck_0/n_0)t\right]\psi(x,t)$ which satisfies the Maxwell equation. 

\begin{figure}[h]
\centering
\includegraphics[width=\linewidth]{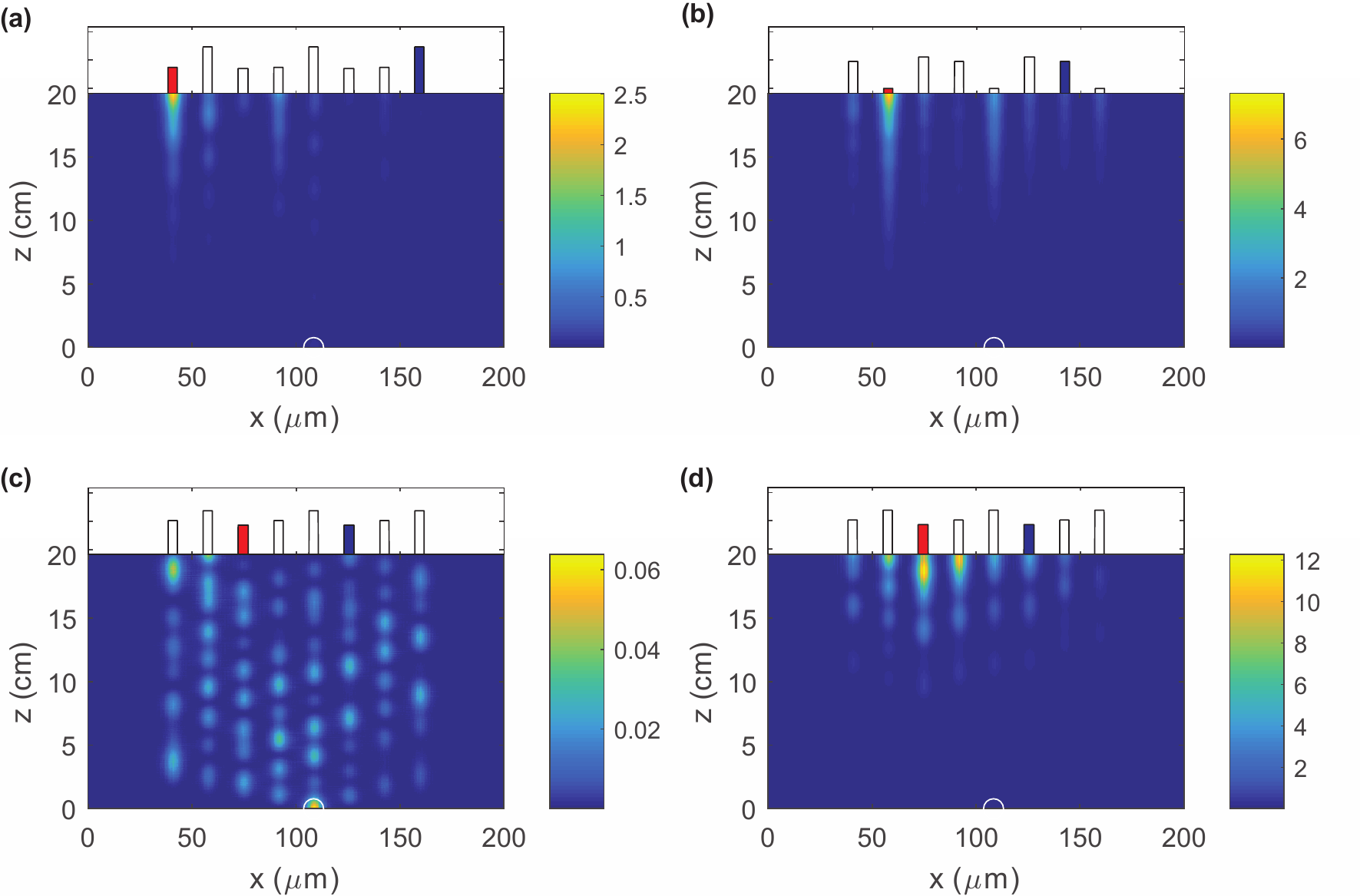}
\caption{BPM simulations of wave packet propagation in an $N=8$ waveguide lattice in the presence of  on-site disorder with period $p=3$. The system parameters are $\lambda$=633 nm, cladding index $n_0=1.45$, waveguide width $W=10$ $\mu$m, and uniform waveguide separation $d=16.9586$ $\mu$m. The bar-chart at the top of each panel shows a random, periodic index-contrast distribution. The vertical scale in each bar-chart denotes the index-contrast $\Delta n$ and ranges from $4.8\mathrm{x}10^{-4}$ to $5.2\mathrm{x}10^{-4}$. {\bf(a)} For a gain potential with strength $\gamma=0.7$ cm$^{-1}$ on the first site, the intensity $I(x,z)$ shows a $\mathcal{PT}$-symmetry broken state. {\bf(b)} With the same gain potential on the second site, the system is again in the $\mathcal{PT}$-broken phase. {\bf(c)} With the same gain-potential on site $m_0=3=p$, the system is in the $\mathcal{PT}$-symmetric phase, as is also shown by site intensities that are one to two orders of magnitude smaller. {\bf(d)} When the gain potential at site $m_0=3$ is doubled to $\gamma=1.4$ cm$^{-1}$, the $\mathcal{PT}$-symmetry is broken. Note that the index-contrast profiles in {\bf(c)} and {\bf(d)} are the same. The BPM analysis confirms the predictions for a nonzero $\mathcal{PT}$-threshold in the presence of random, periodic disorder.}
\label{fig:bpm}
\end{figure}

The index of refraction $n(x)$ is different from that of the cladding only within each waveguide. In the limit of small contrast, $n(x)=n_0+\Delta n$ with $\Delta n/n_0\sim 10^{-4}\ll 1$, the potential term becomes linearly proportional to the index contrast, i.e., $V(x)=2c k_0\Delta n/n_0$, and we implement the gain and loss potentials by adding appropriate imaginary parts to the index contrast. Thus, in the continuum model, a correlated on-site disorder means a random, periodic index-contrast, whereas the tunneling disorder is implemented by random, periodic waveguide separations~\cite{lahini1,bpm3}. Figure~\ref{fig:bpm} shows representative results of such simulations for an $N=8$ waveguide-lattice in the presence of an on-site disorder with period $p=3$. The initial state, marked by a white semicircle, is a normalized Gaussian with width $\sigma=W/2$ in the center waveguide, where $W$ is the width of each waveguide. Each panel shows the time- and space-dependent intensity $I(x,z=ct/n_0)$ where we have switched to the distance along the waveguide $z=ct/n_0$ as a measure of time for an easier comparison with experiments. The bar-chart at the top of each panel shows a randomly generated index contrast $\Delta n(x)$ with period $p=3$. The gain-potential waveguide is shown by a red bar, the reflection-symmetric lossy waveguide is shown by a blue bar, and the linear scale on the vertical axis in each bar-chart ranges from $\Delta n=4.8\mathrm{x}10^{-4}$ to $\Delta n=5.2\mathrm{x}10^{-4}$. 

The intensity plot $I(x,z)$ in Figure~\ref{fig:bpm} (a) is for a gain potential $\gamma=0.7$ cm$^{-1}$ in the first waveguide. It shows that at long times, $z\geq 10$ cm, the intensity is largely confined to the gain waveguide and the system is in the broken $\mathcal{PT}$-symmetry phase. Panel (b) has the intensity plot with the same gain potential, $\gamma=0.7$ cm$^{-1}$, in the second waveguide; it also shows intensity localized in the gain waveguide and thus indicates that the system is in the $\mathcal{PT}$-broken phase. In each case, we note that the maximum intensity $I(x,z)$ is larger than the average intensity $I\sim 1/N=0.125$ expected in each waveguide in the Hermitian limit. Panel (c) shows $I(x,z)$ with the same gain potential strength, $\gamma=0.7$ cm$^{-1}$, in the third waveguide. It is clear from the intensity plot that the system is in the $\mathcal{PT}$-symmetric phase. Note that the gain location $m_0=p=3$ satisfies $N+1=0\mod p$. Panel (d) shows that when the gain potential is doubled, $\gamma=1.4$ cm$^{-1}$, the system enters a $\mathcal{PT}$-broken phase and the resultant intensity is localized largely in the gain waveguide. The results presented in Figure~\ref{fig:bpm} are generic and demonstrate that our findings of a nonzero $\mathcal{PT}$-threshold in disordered lattices are robust (Figure~\ref{fig:ptphase}), as are our predictions for the veiled symmetry of their eigenfunctions (Figure~\ref{fig:parities}). 

\section*{Discussion}

In this paper we have introduced non-Hermitian lattice models with balanced gain and loss that are robust against random, periodic disorder.  We have uncovered a veiled symmetry that is exhibited by eigenfunctions of such disordered, Hermitian lattices. Since this symmetry is phase-sensitive, it ensures equal weights at specific reflections-symmetric sites, but not equal wave functions~\cite{aah}. Therefore, any phase-insensitive observable will reflect the signatures of this symmetry. Experimentally, the models studied here can be realized in coupled waveguide arrays with one gain waveguide and one lossy waveguide. Ideally, if the on-site potentials or tunneling amplitudes are tunable - for example, via voltage-controlled top-gate heaters - it will permit experimental investigations of interplay between localization due to a periodic disorder and the $\mathcal{PT}$-symmetry breaking transition. 

Mathematically, the lattice models considered here correspond to tridiagonal matrices with Hermitian, random, periodic entries, in addition to non-Hermitian, fixed, gain-loss potential entries along the main diagonal. The statistical properties of eigenvalues of such matrices are essentially unexplored. In particular, the dependence of the threshold distribution $PDF(\gamma_{PT})$ on the source and the distribution of disorder is, at this point, poorly understood. A generalization of these models to non-sparse matrices with a positive $\mathcal{PT}$-symmetry breaking threshold, if one were possible, will provide an approach to investigate the spectral properties random, $\mathcal{PT}$-symmetric matrices with real spectra. 


\section*{Acknowledgements}
We thank Tony Lee and Ricardo Decca for insightful comments. This work was supported by NSF Grant no. DMR-1054020. 

\section*{Author contributions statement}
Y.J. conceived the project. A.H. and F.O. carried out numerical calculations. All authors reviewed the manuscript. 

\section*{Additional information}
\textbf{Competing financial interests} The authors declare no competing financial interests. 



\begin{thebibliography}{99}

\bibitem{kubo} Mahan G.D. {\it Many-Particle Physics} (Springer Press, 2000).
\bibitem{keldysh} Kamanev A. {\it Field Theory of Non-Equilibrium Systems} (Cambridge University Press, 2011).  
\bibitem{bender0} Bender C.M. PT symmetry in quantum physics: From a mathematical curiosity to optical experiments. {\it Europhysics News} {\bf 47}, 17-20 (2016). 
\bibitem{review} Joglekar Y.N., Thompson C., Scott D.D., and Vemuri G. Optical waveguide arrays: quantum effects and PT symmetry breaking.  {\it Eur. Phys. J. Appl. Phys.} {\bf 63}, 30001 (2013).
\bibitem{bender1} Bender C.M. and Boettcher S. Real spectra of non-Hermitian Hamiltonians having PT symmetry. {\it Phys. Rev. Lett.} {\bf 80}, 5243 (1998).
\bibitem{bender2} Bender C.M, Brody D.C., and Jones H.F. Complex extension of quantum mechanics. {\it Phys. Rev. Lett.} {\bf 89}, 270401 (2002).
\bibitem{bender3} Bender C.M. Making sense of non-Hermitian Hamiltonians. {\it Rep. Prog. Phys.} {\bf 70}, 947 (2007) and references therein. 

\bibitem{exp1} Guo A. {\it et al.}  Observation of PT-symmetry breaking in complex optical potentials. {\it Phys. Rev. Lett.} {\bf 103}, 093902 (2009). 
\bibitem{exp2} R\"{u}ter C.E., Makris K.G., El-Ganainy R., Christodoulides D.N., Segev M., and Kip D. Observation of parity-time symmetry in optics. {\it Nat. Phys.} {\bf 6}, 192-195 (2010).
\bibitem{exp3} Regensburger A., Bersch C., Miri M.A., Onishchukov G., Christodoulides D.N., and Peschel U.  Parity-time synthetic photonic lattices. {\it Nature} {\bf 488}, 167-171 (2012).
\bibitem{exp4} Peng B. {\it et al.} Parity-time-symmetric whispering-gallery microcavities. {\it Nat. Phys.} {\bf 10}, 394 (2014). 
\bibitem{exp5} Peng B. {\it et al.} Loss induced suppression and revival of lasing. {\it Science} {\bf 346}, 328 (2014).
 \bibitem{bendix} Bendix O., Fleischmann R., Kottos T., and Shapiro B. Exponentially fragile PT-symmetry in lattices with localized eigenmodes. {\it Phys. Rev. Lett.} {\bf 103}, 030402 (2009).
\bibitem{song} Jin L. and Song Z. Solutions of PT-symmetric tight-binding chain and its Hermitian counterpart. {\it Phys. Rev. A} {\bf 80}, 052107 (2009).
\bibitem{znojil1} Znojil M. Gegenbauer-solvable quantum chain model. {\it Phys. Rev. A} {\bf 82}, 052113 (2010).
\bibitem{avadh} Joglekar Y.N. and Saxena A. Robust PT-symmetric chain and properties of its Hermitian counterpart. {\it Phys. Rev. A} {\bf 83}, 050101(R) (2011).
\bibitem{qma} Rubinstein J., Stenberg P., and Ma Q. Bifurcation diagram and pattern formation of phase slip centers in superconducting wires driven with electric current. {\it Phys. Rev. Lett.} {\bf 99}, 167003 (2007). 
\bibitem{serbyn} Serbyn M. and Skvortsov, M.A. Onset of superconductivity in a voltage-biased normal-superconducting-normal microbridge. {\it Phys. Rev. B} {\bf 87}, 020501(R) (2013). 
\bibitem{divergent} Joglekar Y.N., Scott D.D., and Saxena, A. PT-symmetry breaking with divergent potentials: lattice and continuum cases. {\it Phys. Rev. A} {\bf 90}, 032108 (2014). 
\bibitem{moiseyev} Klaiman S., Gunther U., and Moiseyev N. Visualization of branch points in PT-symmetric waveguides. {\it Phys. Rev. Lett.} {\bf 101}, 080402 (2008).

\bibitem{cow1} Christodoulides D.N., Lederer, F., and Silberberg Y. Discretizing light behaviour in linear and nonlinear waveguide lattices. {\it Nature} {\bf 424}, 817-823 (2003).
\bibitem{cow1half} Peschel U., Pertsch T., and Lederer, F. Optical Bloch oscillations in waveguide arrays. {\it Opt. Lett.} {\bf 23}, 1701-1703 (1998).
\bibitem{cow2} Segev, M., Silberberg Y., and Christodoulides, D.N. Anderson localization of light. {\it Nat. Photonics} {\bf 7}, 197-204 (2013). 
\bibitem{bloch} Bloch F. Quantum mechanics of electrons in crystals. {\it Z. Phys.} {\bf 52}, 555-600 (1928). 
\bibitem{anderson} Anderson P.W. Absence of diffusion in certain random lattices. {\it Phys. Rev.} {\bf 109}, 1492-1505 (1958). 
\bibitem{g041} Abrahams, E., Anderson P.W., Licciardello D.C., and Ramakrishnan T.V. Scaling theory of localization: absence of quantum diffusion in two dimensions {\it Phys. Rev. Lett.} {\bf 42}, 673 (1979).
\bibitem{g042} Lee P.A. and Ramakrishnan, T.V. Disordered electronic systems. {\it Rev. Mod. Phys.} {\bf 57}, 287-337 (1985). 
\bibitem{lahini1} Lahini Y. et al. Anderson localization and nonlinearity in one-dimensional disordered photonic lattices. {\it Phys. Rev. Lett.} {\bf 100},  013906 (2008).  
\bibitem{cow3} Mufi A., Anderson localization of light: a tutorial. {\it Adv. Opt. Photonics} {\bf  7}, 459-515 (2015). 
\bibitem{clintdisorder} Thompson C., Joglekar Y.N., and Vemuri G. Disorder effects in tunable waveguides with parity-symmetric tunneling. {\it Phys. Rev. A} {\bf 86}, 043822 (2012). 
\bibitem{molina} Mejia-Cortes C. and Molina M.I. Interplay of disorder and PT-symmetry in one-dimensional optical lattices. {\it Phys. Rev. A} {\bf 91}, 033815 (2015). 
\bibitem{konotop1} Khartashov Y.V., Hang C., Konotop V.V., Vysloukh V.A., Huang G., and Torner L. Suppression and restoration of disorder-induced light localization mediated by PT-symmetry breaking. {Laser \& Photon. Rev.} {\bf 10}, 100-107 (2016). 

\bibitem{berry} Bender C.M., Berry M.V., and Mandilara A. Generalized PT symmetry and real spectra. {\it J. Phys. A: Math. Gen.} {\bf 35}, L467-L471 (2002). 
\bibitem{ali} Mostafazadeh A. Pseudo-Hermiticity versus PT symmetry: the necessary condition for the reality of the spectrum of a non-Hermitian Hamiltonian. {\it J. Math. Phys.} {\bf 43}, 205-214 (2002). 
\bibitem{mark} Joglekar Y.N., Scott D.D., Babbey M., and Saxena A.  Robust and fragile PT-symmetric phases in a tight-binding chain. {\it Phys. Rev. A} {\bf 82}, 030103 (R) (2010).  
\bibitem{derekring} Scott D.D. and Joglekar Y.N. PT-symmetry breaking and ubiquitous maximal chirality in a PT-symmetric ring. {\it Phys. Rev. A} {\bf 85}, 062105 (2012). 
\bibitem{aah} Harter A.K., Lee T.E., and Joglekar Y.N. PT-breaking threshold in spatially asymmetric Aubre-Andre and Harper models: hidden symmetry and topological states. {\it Phys. Rev. A} {\bf 93}, 062101 (2016).

\bibitem{rmt} Mehta M.L. {\it Random matrices} (Academic Press, 2004). 
\bibitem{pd1} Izrailev F.M. and Krokhin A.A. Localization and the mobility edge in one-dimensional potentials with correlated disorder. {\it Phys. Rev. Lett.} {\bf 82}, 4062-4065 (1999). 
\bibitem{pd2} Lazo E. and Onell M.E. Extended states in 1-D Anderson chain diluted by periodic disorder. {\it Physica B: Cond. Matter} {\bf 299}, 173-179 (2001). 

\bibitem{bpm3} Szameit A. and Nolte S. Discrete optics in femtosecond-laser-written photonic structures. {\it J. Phys. B: At. Mol. Opt. Phys.} {\bf 43}, 163001 (2010). 
\bibitem{bpm1} Chung Y. and Dagli N. An assessment of finite difference beam propagation method. {\it IEEE J. Quant. Electron.} {\bf 26}, 1335-1339 (1990). 
\bibitem{bpm2} Shakir S.A., Motes R.A., and Berdine R.W. Efficient scalar beam propagation method. {\it IEEE J. Quant. Electron.} {\bf 47}, 486-491 (2011).

\end{thebibliography}
\end{document}